# Mixed order parameter symmetries in cuprate superconductors

A. Bussmann-Holder[1], R. Khasanov[2], A. Shengelaya[3], A. Maisuradze[2], F. La Mattina[2], H. Keller[2] and K. A. Müller[2]

[1]*Max- Planck-Institut für Festkörperforschung, Heisenbergstr. 1, D-70569 Stuttgart, Germany*

[2]*Physik Institut der Universität Zürich, Winterthurerstrasse 190, CH-8057 Zürich, Switzerland*

[3]*Physics Institute of Tbilisi State University, Chavchadze 3, GE-0128 Tbilisi, Georgia*

The recent observation of an inflection point in the temperature dependence of the in-plane magnetic field dependence ($\lambda_{ab}$) is investigated within a two-band model with coupled order parameters of different symmetries. While the dominant order parameter has d-wave symmetry, the smaller one is of s-wave symmetry. Superconductivity is robust in the d-wave channel and induced via interband interactions in the s-wave subsystem.

Even though various experiments reveal that the leading order parameter in cuprate superconductors is of d-wave symmetry, there are a variety of other experiments which show that an additional order parameter symmetry must be present as well [1]. These findings are not at all surprising since nearly all cuprate superconductors are orthorhombic which is incompatible with a pure d-wave pairing potential. Also in recent tunnelling experiments subgap features have been observed which have been interpreted in terms of coupled order parameters of s+id symmetry [2, 3]. Early on, it has already been suggested [1] that obviously controversial observations of the superconducting order parameter symmetry could be resolved by assuming that both, s- and d-wave symmetries are realized simultaneously in cuprates. This has, however, not been pursued very much in the following, partially due to the fact that coupled multiple order parameters have been considered to be an exception in superconductivity. With the discovery of $MgB_2$ and the unambiguous finding of two coupled order parameters of s+s symmetry [4] the situation has changed completely and various other superconductors, including heavy fermion compounds, have been shown to exhibit multiple coupled gaps [5].

Recent muon-spin rotation (μSR) studies of single-crystal $La_{1.83}Sr_{0.17}CuO_4$ [6] revealed the presence of an inflection point in the temperature dependence of the in-plane magnetic penetration depth which was also observed in $MgB_2$ [7]. The observation of this inflection point is a clear signature that two gaps are present with one gap being the leading one whereas the second has to be substantially smaller in magnitude than the leading one. Only such a strong difference in the gap values is able to produce these features in the temperature dependence of $\lambda_{ab}$.

In a variety of previous work we have already addressed the issue of multiple gaps theoretically since the complexity of the Fermi surface in cuprates cannot be accounted for by a one-band-only approach [8]. Similar conclusions have been reached by various groups before [9], and special emphasis has been dedicated to the fact, that within multi-gap superconductors enormous enhancements of the superconducting transition temperature can be achieved even within weak coupling approaches [8]. Since the derivation of the gap equations has been presented in detail in previous work [8], we concentrate here on the calculation of the squared inverse penetration depth $1/(\lambda_{ab})^2$ using a two band model. The normalized quantity $\rho(T) = \lambda_{ab}^2(0)/\lambda_{ab}^2(T)$ is given by:

$$\rho(T) = 1 + 2\sum_i \int_{\Delta_i(T)}^{\infty} \left(\frac{\partial f}{\partial E}\right) \frac{E}{\sqrt{E^2 + \Delta_i(T)^2}} dE \qquad (1)$$

where $f = 1/[1+\exp(E/k_B T)]$ is the Fermi function and the summation is over $i = 1,2$. The momentum *k* dependent band energies are given by:

$$E(k_i) = 2t_1(\cos k_x a + \cos k_y b) - 4t_2(\cos k_x a \cos k_y b) - \mu \qquad (2)$$

with *a, b* being the lattice constants in the $CuO_2$ planes, and $a \neq b$ to account for the orthorhombicity, and $\mu$ is the chemical potential which controls the doping. In analogy with previous work we incorporate explicitly the coupling to the lattice through polaron formation [10]. This renormalizes the band dispersion in two ways by first leading to a rigid band shift, and second by inducing an exponential band narrowing. These renormalizations have been shown to be of extreme importance in understanding the unconventional isotope effects observed in cuprates [10, 11].

We assume here that two gaps contribute to $\rho$ where one gap has d-wave symmetry ($\Delta_d$) whereas the other one is isotropic s-wave ($\Delta_s$). It is important to note, that both gaps are not independent of each other but are coupled through interband interactions with each other like:

$$\Delta_d = \Delta_{k_1} = \sum_{k_1'} V_1(k_1, k_1')\Delta_{k_1'}\Phi_{k_1'} + \sum_{k_2} V_{1,2}(k_1, k_2)\Delta_{k_2}\Phi_{k_2} \qquad (3a)$$

$$\Delta_s = \Delta_{k_2} = \sum_{k_2'} V_2(k_2, k_2')\Delta_{k_2'}\Phi_{k_2'} + \sum_{k_1} V_{1,2}(k_1, k_2)\Delta_{k_1}\Phi_{k_1} \qquad (3b)$$

where the $V_i$ (*i*=1,2) are effective attractive intraband pairing potentials, whereas $V_{12}$ is the interband pairing potential which induces pairwise exchange between the two bands. The temperature dependencies of the gaps are given by $\Phi_{k_i} = \frac{1}{2\xi(k_i)}\tanh\left[\frac{\xi(k_i)}{2k_B T}\right]$, and $\xi(k_i) = \sqrt{E(k_i)^2 + \Delta_{k_i}^2}$. The coupled gaps (eqs. 3a, 3b) are calculated simultaneously and self-consistently as a function of temperature with the interaction potentials chosen such as to fit the experimental values of the zero temperature gaps and $T_c$ [6]. Note, however, that the interaction potential in the s-wave channel, $V_2$, is negligibly small and superconductivity is induced here through the interband coupling term $V_{12}$. This means that two free parameters are used to calculate the temperature dependent gaps, $T_c$ and the penetration depth. The calculated gaps at zero magnetic field are shown in fig. 1.

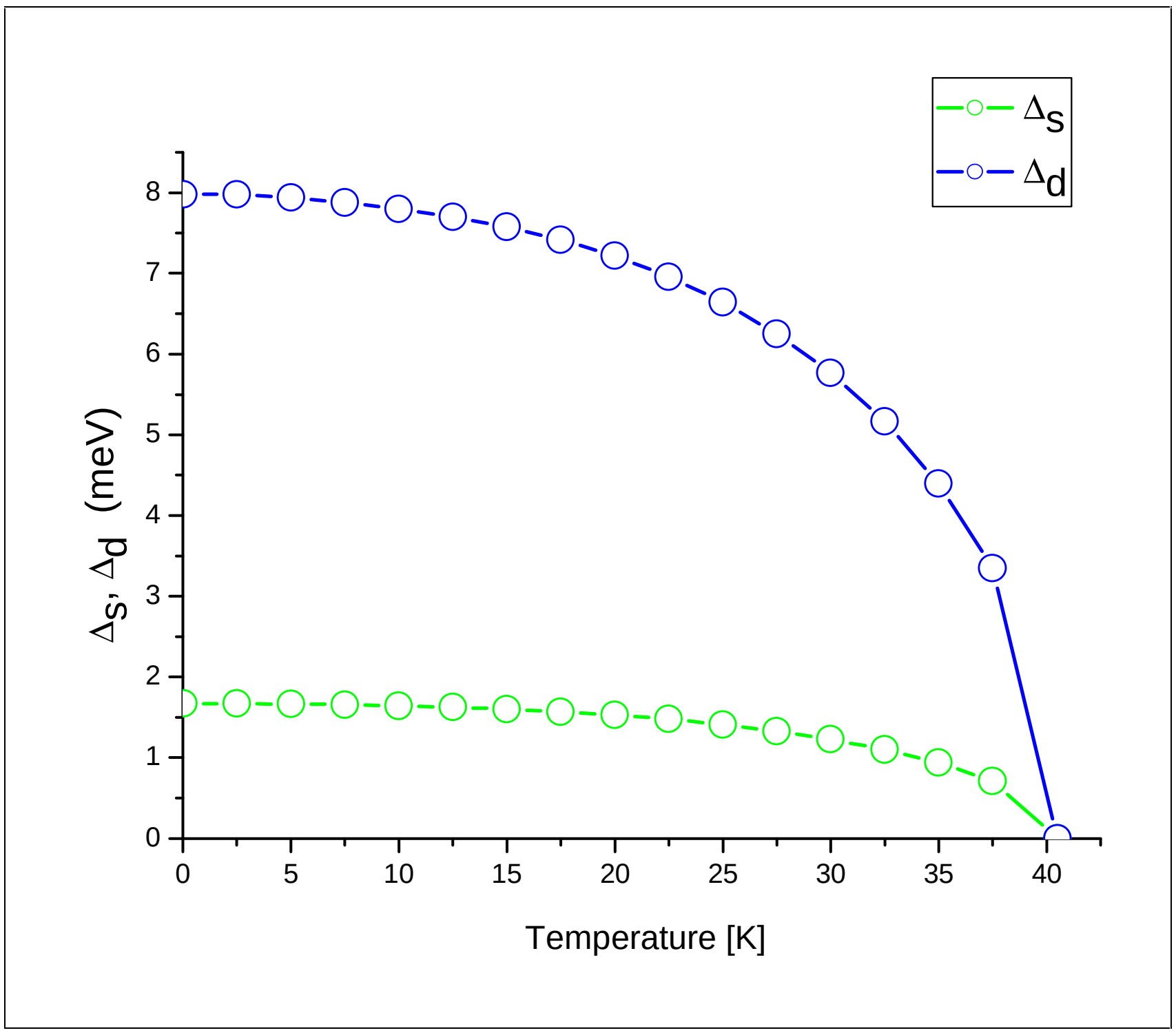

**Figure 1** Calculated temperature dependencies of $\Delta_d$ and $\Delta_s$ (eqs. 3a and 3b) which have been used to calculate $\rho(T)$ using eq. 1.

The experimentally determined strong anisotropy of the gaps is well reproduced. While the larger gap $\Delta_d$ with d-wave symmetry is strongly temperature dependent, the smaller one $\Delta$ with s-wave symmetry is nearly temperature independent over a large temperature scale. A very similar behaviour has also been observed in $MgB_2$ [4]. It is important to emphasize that due to the gap coupling both gaps loose their individual character as arising from their specific symmetry and adopt a strong component from the other gap symmetry. This has important implications since d-wave specific properties as, e.g., linear in T dependence of the penetration depth at low temperatures, is not present. Interestingly, also experimental data confirm these results [6]. The superfluid density is calculated with eq. 1 and using similar weights for the two gaps contributing to $\rho$ as the experimentally determined ones [6]. The effect of the magnetic field on $\rho$ has been shown to lead to a small reduction in $T_c$ and the two gaps. Within the calculation this is simulated by changing the interband interaction potential which is crucial in determining the absolute value of $T_c$. The theoretical results for $\rho$ for different magnetic fields are compared to the experimental ones in fig. 2.

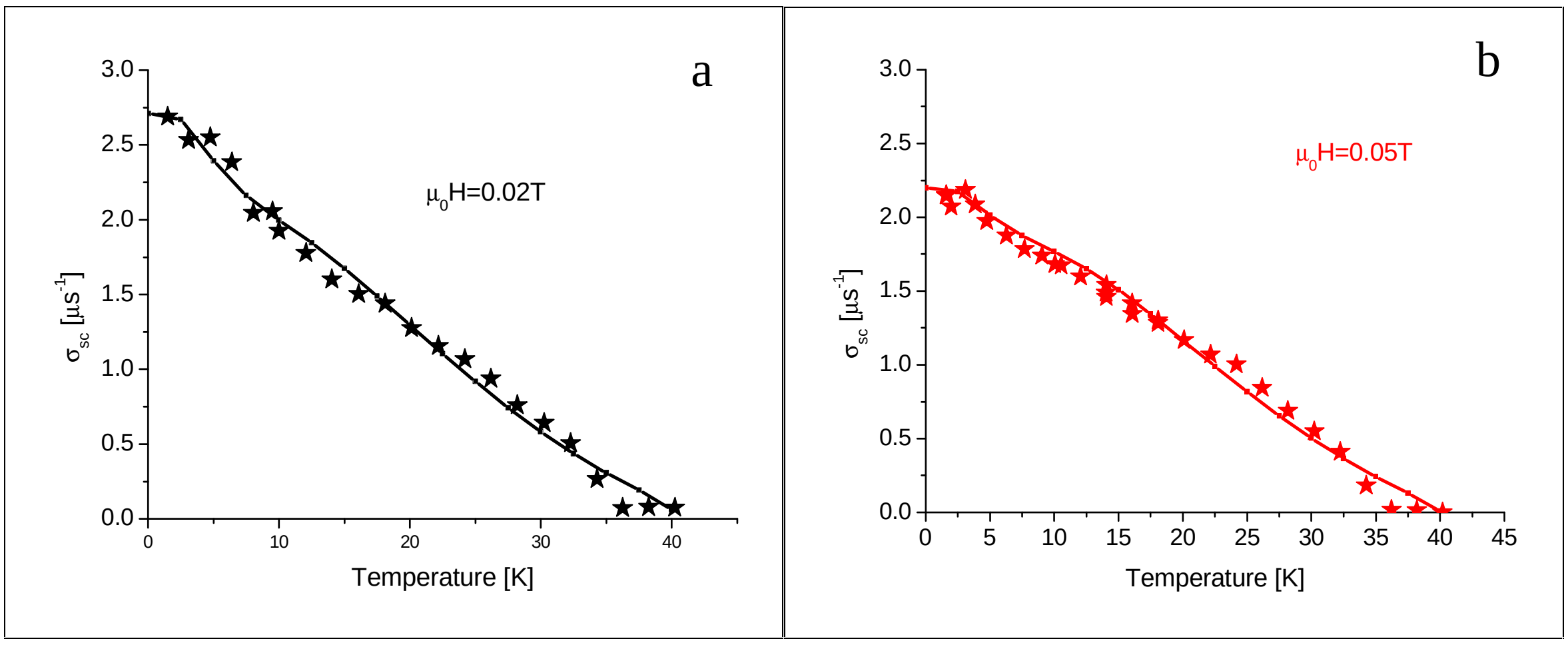

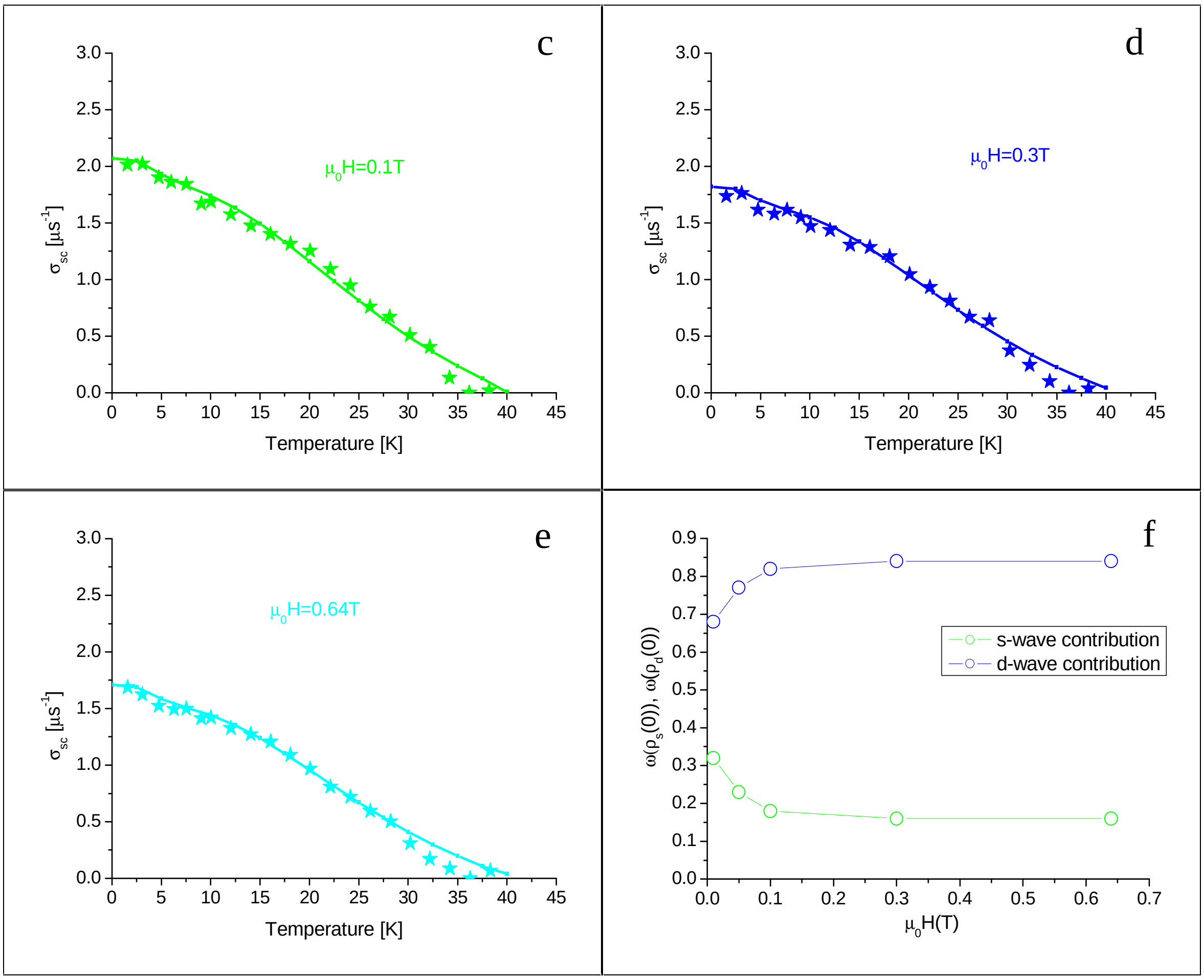

**Figure 2 (a – e)** Comparison between the experimental μSR relaxation data $\sigma_{sc}(T) \propto 1/\lambda_{ab}^2(T)$ for single crystal $La_{1.83}Sr_{0.17}CuO_4$ (stars) [6] and theoretical results for $\sigma_{sc}(T) \propto \sigma_{sc}(T)$ (full lines) for various magnetic fields $\mu_0 H$.

**Figure 2f** The calculated individual contributions $\omega_d(0)$ and $\omega_s(0) = 1 - \omega_d(0)$ to the total $\rho$ as functions of $\mu_0 H$. Here $\rho(T) = \omega_d \rho_d(T) + \omega_s \rho_s(T)$.

together with the individual contributions $\rho_s$ and $\rho_d$ to $\rho$ from the two gaps. While the agreement with the experimental data is overall very good, discrepancies are observed for temperatures close to $T_c$ where the theoretical results overestimate $\rho$. The individual contributions from the $\rho_d$ and $\rho_s$ (fig. 2f) are essentially identical to those obtained from the experiment. As can be seen there, there is only a minor field dependence in the individual contributions to $\rho$, which is completely lost for fields larger than 0.15 T. This implies that both $\rho_d$ and $\rho_s$ are suppressed by the magnetic field in the same manner. Interestingly, also the anisotropy in the Cu NMR relaxation time shows the same fraction of s+d admixture, i.e., 20% s-wave and 80% d-wave, consistent with the results for $\rho$ [12]. The unusual appearance of an inflection point observed in all $\sigma_{sc}(T)$ data in figure 2 is also a consequence of this admixture since equal contributions from both components would suppress the inflection point. The temperature dependent individual contributions are shown in Fig. 3 in order to elucidate the origin of this inflection point.

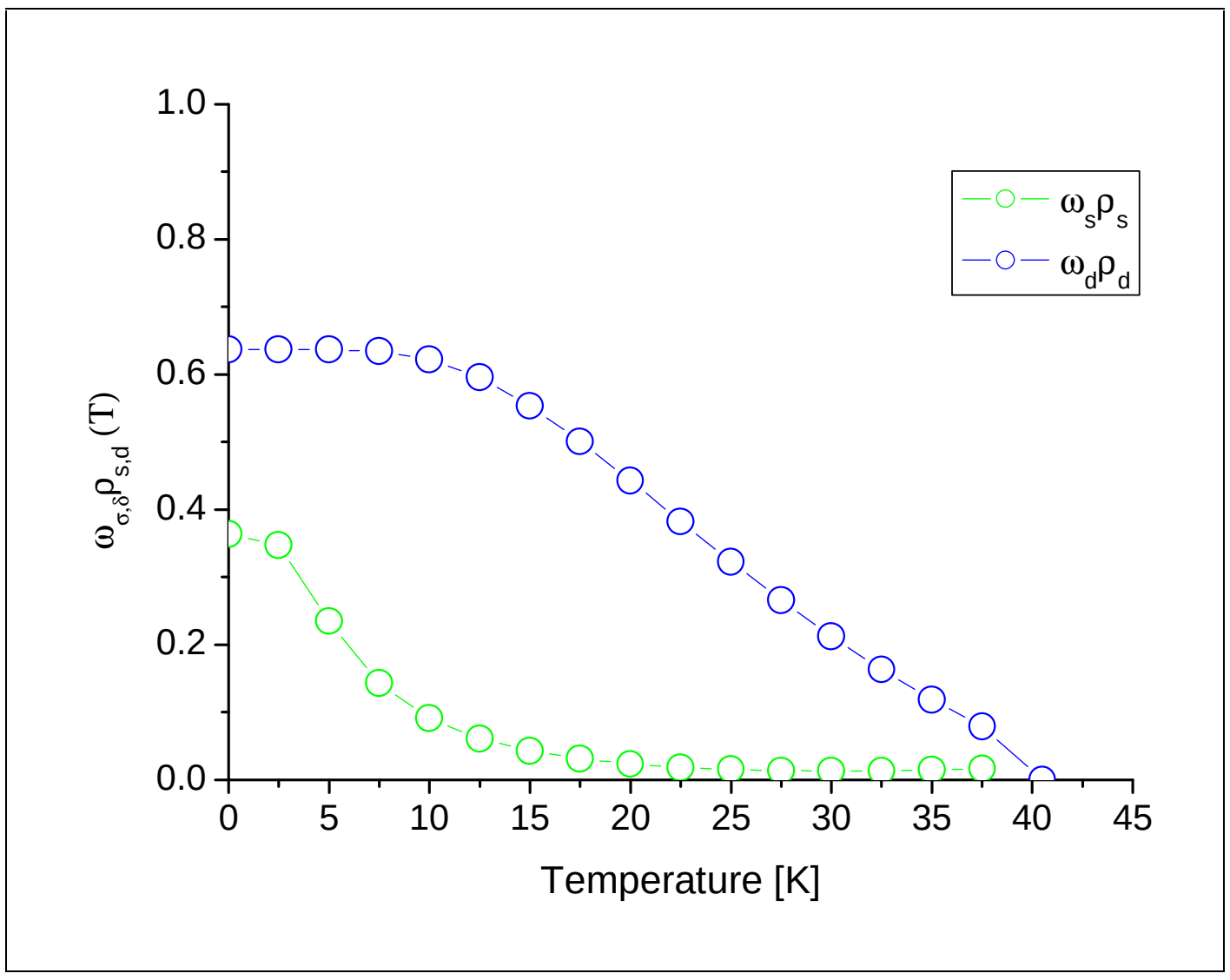

**Figure 3** Calculated temperature dependencies of the weighted contributions of $\omega_s \rho_s(T)$ and $\omega_d \rho_d(T)$ (see Fig. 2f) with $\omega_d + \omega_s = 1$.

While the larger component $\rho_d$ shows a conventional temperature dependence (note, that the linear in T low temperature behaviour is absent for the d-wave gap due to s-wave admixture), the small component $\rho_s$ rapidly decreases with increasing T to have nearly no contributions already at $T_c/2$.

As has been pointed out above the magnetic field dependence of the data has been incorporated on a phenomenological level only, by scaling the interband pair potential to fit the experimental data. Here it is assumed that a linear relation between $V_{12}$ and $\mu_0 H$ is given where $V_{12}$ decreases with increasing $\mu_0 H$ like $V_{12}(\mu_0 H) = 1.3885 - 0.2992\mu_0 H$.

Even though our choice of combining s+d wave order parameters to calculate the penetration depth might seem to be rather arbitrary since other possibilities could, e.g., be d+d, d+is etc. This choice can be justified by analyzing the field dependence of the zero temperature data $\rho_s(0), \rho_d(0)$. This is shown in fig. 4 where the following linear relations are obtained: $\rho_d(0) \approx 0.65 - 0.2\sqrt{\mu_0 H}$ whereas $\rho_s(0) \approx 0.07 + 0.03/\sqrt{\mu_0 H}$, both results are in agreement with theoretical results for a d-wave, and s-wave gap function [13].

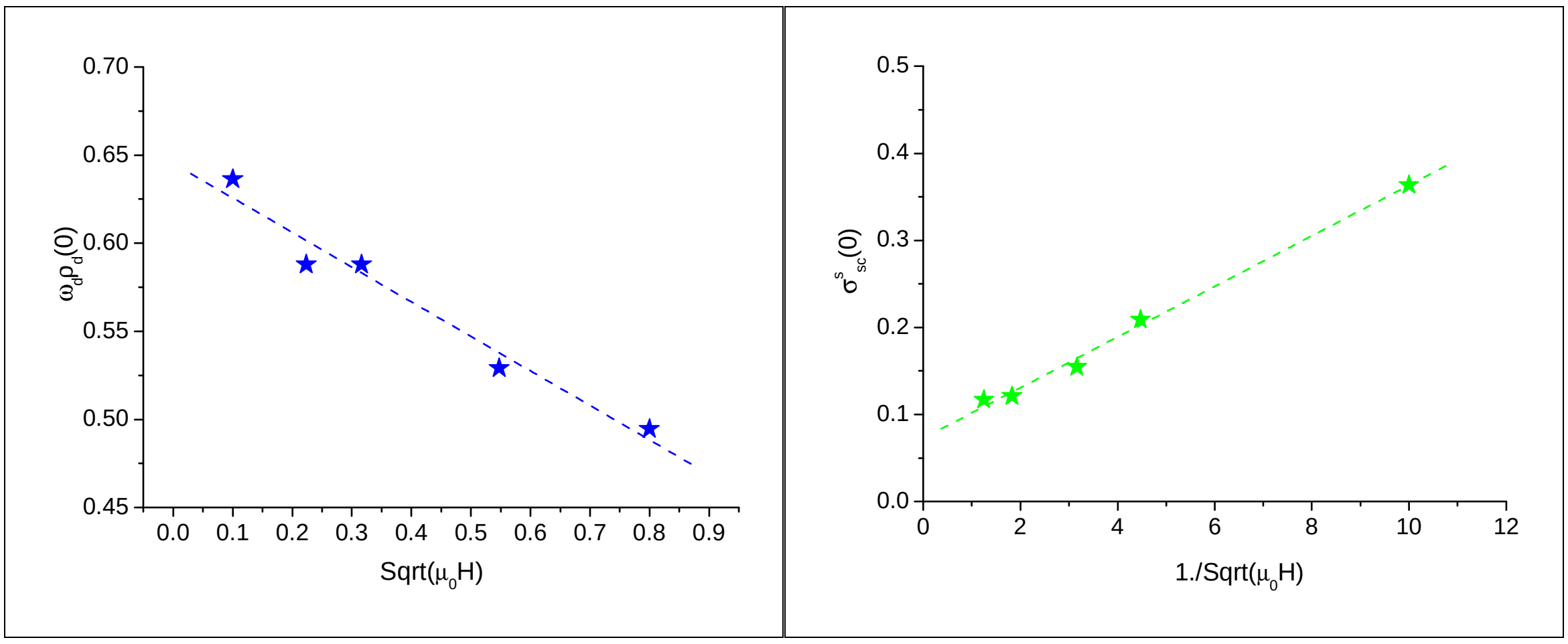

**Figure 4** Magnetic field dependencies of the zero temperature penetration depth for the d-wave component $\omega_d \rho_d(0)$ (left panel) and the s-wave component $\omega_s \rho_s(0)$ (right panel). Note that both data sets have been normalized such that the zero field components add to unity. The linear dependence of $\omega_s \rho_s(0)$ on $1/\sqrt{\mu_0 H}$ is obviously not valid for too large fields.

In conclusion, we have analyzed the experimentally observed temperature dependence of the in-plane penetration depth $\lambda_{ab}$ within a two-band model with interband interactions. Effectively attractive intraband interactions with d- and s-wave symmetry have been inferred together with an interband scattering potential where the interaction in the s-wave channel is too small to induce superconductivity here. However, the interband interaction supports superconductivity also in this band and leads to the appearance of a rather small superconducting gap as compared to the leading d-wave gap. The unusual observation of an inflection point in the temperature dependence of $\lambda_{ab}$ is a direct consequence of the large discrepancy in the two gap values. The choice of combining a d-wave order parameter with an s-wave one is justified by the analysis of the field dependencies of the zero temperature values of the two contributions to $\rho(0)$ which show the expected behaviour as functions of $\mu_0 H$. Finally, it is important to mention that it is impossible to use a single component approach to high $T_c$ cuprate superconductors in understanding these new features.

**Acknowledgement** This work was supported by the Swiss National Science Foundation, the K. Alex Müller Foundation, the EU project CoMePHS, and in part by the NCCR program MaNEP.